\documentclass[12pt]{article}
\usepackage{amsmath}
\usepackage{amsfonts}
\usepackage{amssymb}
\usepackage{xcolor}
\usepackage{color}
\usepackage[english]{babel}
\usepackage{graphics}
\usepackage{graphicx}
\begin{document}
\title{ Comments on\\ "Which is the Quantum Decay Law\\ of
Relativistic Particles?"}
\author{K. Urbanowski\footnote{e--mail: K.Urbanowski@proton.if.uz.zgora.pl},\\
\hfill\\
University of Zielona G\'{o}ra, Institute of Physics, \\
ul. Prof. Z. Szafrana 4a, 65--516 Zielona G\'{o}ra, Poland.\\
}
\maketitle
\abstract{
Results presented in a recent paper {\em "Which is the Quantum Decay Law of
Relativistic particles?"}, arXiv: 1412.3346v2 [quant--ph]], are analyzed.
We show that approximations used therein to derive the main final formula for
the survival probability of finding a moving  unstable particle to be undecayed at time $t$
force this particle  to almost stop moving, that is that, in fact, the derived formula is approximately
valid only for $\gamma \cong 1$, where $\gamma = 1/\sqrt{1-\beta^{2}}$
and $\beta = v/c$, or in other words, for the velocity  $v \simeq 0$.
}
\hfill\\
\hfill\\
\hfill

In a recent paper \cite{carlo} authors  analyze the relation between the quantum mechanical
survival probability of the moving unstable particle and that of the particle at rest.
Based on the results of this analysis they call in question conclusions and results
presented in \cite{stefanovich,stefanovich1,stefanovich2,shirkov,ku-plb}.
In these papers it has been shown that rigorous quantum mechanical treatment of the problem
leads to the conclusion that
in the case of quantum decay processes
the standard, classical time dilation
formula for the survival probabilities
and the standard relation between the decay lifetime of the
moving relativistic particle and the particle in the rest
work in good approximation for times no longer than a few lifetimes
(which is in full agreement with the known experimental results).
It is also shown therein
that for times much longer than a few lifetimes
it may not work.

In \cite {ku-plb} the decay curves of the moving relativistic unstable particles has been
presented. They were
found numerically for particles modeled by Breit--Wigner mass distribution
starting from the instant of the creation of the particle up to
100 lifetimes and even 1000 lifetimes depending on the parameters
of the model considered. These results confirm
the analytical result presented in \cite{stefanovich,stefanovich1,stefanovich2,shirkov}
and discussed also in \cite{ku-plb}.

Note that,  the classical  time dilation relation has been verified in many experiments
as it was  also stated in \cite{carlo}, but
in all known experiments this relation was verified
for times of order of the lifetime, and generally for times
no longer than a few lifetimes \cite{muon1,muon2}.
There is not any experimental confirmation of this relation for
times longer than a few lifetimes and for much longer times, e.g. for times
when the late time deviations of the decay law from the exponential form
begin to dominate. So, taking into account these facts
one can not state that the conclusion drawn in
\cite{stefanovich,stefanovich1,stefanovich2,shirkov,ku-plb} that the
classical dilation formula sufficiently well reflects experimental data only
for times no longer than a few lifetimes,  is in conflict
with results of the experiments mentioned.

There are some statements in \cite{carlo} which are not obvious, but
this does not affect  the calculations performed therein.
So, let us pass to the analysis of the calculations leading to the
final conclusion of \cite{carlo}.

One of the main differences between approach used in \cite{carlo} and that
used in \cite{stefanovich,stefanovich1,stefanovich2,shirkov,ku-plb} is that within
the quantum mechanical treatment of the problem
authors of \cite{carlo} consider  the  probability amplitude of the survival
probability assuming that moving particles have a definite velocity $\vec{v}$ whereas
authors of \cite{stefanovich,stefanovich1,stefanovich2,shirkov,ku-plb} assume
that the momentum $\vec{p}$ of the particles considered is definite.
The assumption that the momentum is definite allows one
to avoid  inconsistences caused by the assumption that the velocity is
definite (see, eg. (19), (20) in \cite{carlo} or discussion
presented in \cite{shirkov1,shirkov2}). Moreover, such an assumption is based on
the first principles:
According to the fundamental principles of the classical physics and quantum
theory (including relativistic quantum field
theory) the energy and momentum of the
moving particle have to be conserved.
These conservation laws are one of the
basic and  model independent tools of the study
of reactions between the colliding or decaying particles.
There is no an analogous conservation law for the velocity $\vec{v}$.
Therefore it seems to be  more reasonable to assume
that momentum $\vec{p}$ of the moving unstable particles measured
in the rest frame of  the observer is definite and constant.

 In \cite{stefanovich,stefanovich1,stefanovich2,shirkov,ku-plb} the case of the
 moving particle observed in the rest frame of the observer is analyzed: In such
 a case states $|\Phi_{p}\rangle$, (where $|\Phi_{p}\rangle$ is the the state vector
 of the moving unstable particle with the definite momentum $\vec{p}$, which is measured
 by the observer in his rest reference frame),  and
$|\Phi_{0}\rangle \equiv |\Phi_{p=0}\rangle$, are different states
belonging  to the same Hilbert space ${\cal H}_{0}$ connected with the rest frame of the observer.
The corresponding survival probabilities are defined as follows:
$P_{0}(t) = |A_{0}(t)|^{2}$ and $P_{p}(t)= |A_{p}(t)|^{2}$, where
\begin{eqnarray}
A_{0}(t) &=& \langle \Phi_{0}|e^{\textstyle{-itH}}|\Phi_{0}\rangle, \label{P0}\\
A_{p}(t) &=& \langle \Phi_{p}|e^{\textstyle{-itH}}|\Phi_{p}\rangle, \label{Pp}
\end{eqnarray}
and $H$ is the total selfadjoint Hamiltonian of the system considered. (The system of units
$\hbar = c = 1$ is used).

In \cite{carlo} the final result is obtained for states
connected with the {\em "reference frame in which the
system is in motion with velocity $\vec{v}$"}. In this new reference
frame the momentum of the particle equals $\vec{k}_{m}$ and $\vec{k}_{m} \neq \vec{p}$,
where $\vec{p}$ is the momentum of the same particle but in the rest frame
of the observer.
The state of the moving unstable particle is described by
a vector $|\Phi_{v}\rangle$   which should be an element of
the Hilbert space ${\cal H}_{v}$
connected with this new reference frame in which the system is in motion
but this problem is not explained in \cite{carlo}.
So, in fact, the case considered in  \cite{carlo} is not the same case,
which was analyzed  in \cite{stefanovich,stefanovich1,stefanovich2,shirkov,ku-plb}.
What is more, in \cite{carlo} the final result has been obtained for
the amplitude (see (21) therein),
\begin{equation}
A_{v}(t;\vec{x}) = \langle \Phi_{v}|e^{\textstyle{-itH +
i\vec{P}\cdot \vec{x}}}|\Phi_{v}\rangle, \label{Avx1}
\end{equation}
where $\vec{x}$ is a coordinate.
This is a fact. So, comparing with each other the  results reported in
\cite{carlo} and presented in \cite{stefanovich,stefanovich1,stefanovich2,shirkov,ku-plb}
is at least doubtful. What is more,
an interpretation of the amplitude $A_{v}(t;\vec{x})$ is unclear:
The vector $\exp\,[-itH + i \vec{P}\cdot \vec{x}]\,|\Phi_{v}\rangle$ does not
solve the evolution equation for the initial condition $|\Phi_{v}\rangle$.

Searching  for the properties of the amplitude  $A_{v}(t; \vec{x})$
authors of \cite{carlo} (as well as authors of
\cite{stefanovich,stefanovich1,stefanovich2,shirkov,ku-plb}
analyzing properties of $A_{0}(t)$ and $A_{p}(t)$) use
the integral representation of $A_{v}(t; \vec{x})$ as the
Fourier transform of the energy or,
equivalently mass distribution function $\omega (m)$
(see, eg. \cite{Fock,khalfin,fonda,muga}) and obtain that (see (39) in \cite{carlo})
\begin{eqnarray}
A_{v}(t;\vec{x}) &=&
\int \, dm\, \Big[\omega(m) \times \label{Avx2} \\
&& \times
\int\,d^{3}\vec{p}\,|\phi (\vec{p})|^{2}\,
e^{\textstyle{-iE_{m}(\vec{k}_{m})\,t\,+\,i\vec{k}_{m}\cdot \vec{x}}}\;\Big], \nonumber
\end{eqnarray}
where $\omega (m) = |\rho (m)|^{2}$ and $\rho (m)$ are the expansion
coefficients of $|\Phi_{v}\rangle$ in the basis of eigenvectors
$|E_{m}(\vec{k}_{m}), \vec{k}_{m},m\rangle$ for the Hamiltonian $H$
(see (37) in \cite{carlo}).  $\phi (\vec{p})$ is the momentum distribution
such that $\int d^{3}\vec{p}\,|\phi (\vec{p})|^{2} =1$. The energy
$E_{m}(\vec{k}_{m})$ and momentum $\vec{k}_{m}$ in the new reference
frame mentioned are connected with $E_{m}(\vec{p})$ and $\vec{p}$ in the
rest frame by Lorentz transformations (see (33) --- (35) in \cite{carlo}),
\begin{eqnarray}
E_{m}(\vec{k}_{m}) &=& \gamma (E_{m}(\vec{p}) + v \, p_{\parallel}), \label{Ekm} \\
k_{m\,\parallel} & = & \gamma (p_{\parallel} + v E_{m}(\vec{p}), \label{km} \\
\vec{k}_{m\,\perp} &=& \vec{p}_{\perp}, \label{k-perp}
\end{eqnarray}
where $k_{m\,\parallel} \; (\vec{k}_{m\,\perp})$ and
$p_{\parallel}\; (\vec{p}_{\perp})$ are components of
$\vec{k}_{m}$ and $\vec{p}$ parallel (orthogonal)
to the velocity $\vec{v}$, and
\begin{equation}
E_{m}(\vec{p}) = \sqrt{m^{2} + \vec{p}^{\;2}}. \label{Emp}
\end{equation}
Next, authors of \cite{carlo} limited their considerations to the case when
for the decay width $\Gamma$,  mass of the particle $M$ and the  momentum
uncertainty $\sigma_{p}^{2} = \int \, d^{3}\vec{p}\,|\phi (\vec{p})|^{2}(p_{i})^{2}$, ($i=1,2,3$),
the following condition
\begin{equation}
\Gamma  \ll \sigma_{p} \ll M, \label{sigma}
\end{equation}
is assumed to hold. This is crucial condition which allowed them
to approximate the energy $E_{m}(p)$ as follows
\begin{equation}
E_{m}(\vec{p})
\simeq m, \label{Emp-1}
\end{equation}
neglecting terms of order $\vec{p}^{\;2}/m^{2}$. A discussion
of the admissibility of the condition (\ref{sigma})
uses arguments similar to those one can find, e.g. in \cite{exner}.
The difference is that in \cite{exner} the approximation
$E_{p}(m) \simeq m + \vec{p}^{\;2}/2m$ is used instead of (\ref{Emp-1}).

The condition (\ref{sigma}) and its consequence, that is the approximation (\ref{Emp-1}),
were used in \cite{carlo} to replace  relations (\ref{Ekm}), (\ref{km})
by the following, approximate relations,
\begin{eqnarray}
E_{m}(\vec{k}_{m}) &\simeq& \gamma (m + v \, p_{\parallel}), \label{Ekm-a} \\
k_{m\,\parallel} & \simeq & \gamma (p_{\parallel} + v m). \label{km-a}
\end{eqnarray}
Finally replacing $E_{m}(\vec{k}_{m})$ and $\vec{k}_{m}$ under the integral sign
in (\ref{Avx2}) by (\ref{Ekm-a}) and (\ref{km-a}) respectively (or in \cite{carlo},
in (41) by (33) and (34))
after some algebra authors of \cite{carlo} obtain the  relation (46) that was needed, that is
a relation of the following type
\begin{equation}
P_{v}(t) = P_{0}(t/\gamma). \label{P=Pv}
\end{equation}
This result obtained within the condition and approximation described
above was the basis of the all conclusions presented in \cite{carlo}.

Unfortunately, in \cite{carlo} only the admissibility of the assumed
conditions and approximations used was discussed
without any analysis of their physical consequences.
Let us note that assuming (\ref{Emp-1}) and keeping the term $
v p_{\,\parallel} \neq 0$ in $(\ref{Ekm-a})$ is justified only if the contribution of
$v p_{\,\parallel}$ into (\ref{Ekm-a}) is not negligible, ie., when
$|v p_{\,\parallel}| > 0$ significantly: It is  because within the system of units used
$v$ is dimensionless quantity such that $0\leq v < 1$ and thus always
$|v p_{\,\parallel} < |p_{\,\parallel}|$. This means that in fact the
contribution of $v p_{\,\parallel}$ is not negligible
if $|p_{\,\parallel}| > 0$ (and thus $|\vec{p}|>0$) significantly.
In such a case the use of the identity
\begin{equation}
\gamma \equiv \frac{E_{m}(\vec{p})}{m} = \frac{\sqrt{m^{2} + \vec{p}^{\;2}}}{m}. \label{gamma-p}
\end{equation}
to find that in the case of (\ref{Emp-1}) the conclusion
\begin{equation}
\gamma \equiv \frac{E_{m}(\vec{p})}{m} \simeq 1, \label{gamma=1}
\end{equation}
seems to be  justified and it will be reasonable as long as
the nonzero term $vp_{\,\parallel}$ will be used in (\ref{Ekm-a}).

The similar analysis performed within the use of the identity (\ref{gamma-p}) shows that the approximation used in
\cite{exner} leads to the conclusion that the results obtained therein are valid for
$\gamma \simeq 1 + \vec{p}^{\;2}/2m^{2} < 2$ , that is for
$1 < \gamma <  2$, and for no more than a few lifetimes.

We should remember that $E_{m}(\vec{p})$ and $\vec{p}$ used in (\ref{Ekm}), (\ref{km})
and (\ref{Ekm-a}), (\ref{km-a})   can not be considered as they would be completely
independent quantities. (In fact, it was done in \cite{carlo} using
the approximations (\ref{Ekm-a}), (\ref{km-a})).
The supposition, that the contribution of $v p_{\,\parallel}$ into (\ref{Ekm-a})
can not be neglected, contradicts the approximation used to obtain the result
(\ref{Emp-1}) which is crucial for the main conclusions of \cite{carlo}. Simply
\begin{equation}
E_{m}(\vec{p}) = \sqrt{m^{2} + \vec{p}^{\;2}} \simeq m \;\;\Leftrightarrow\;\;|\vec{p}| \simeq 0, \label{Emp-p1}
\end{equation}
or, more exactly,
\begin{equation}
E_{m}(\vec{p}) = \sqrt{m^{2} + \vec{p}^{\;2}}
\simeq m \;\;\Leftrightarrow\;\;|\vec{p}_{\perp}| \simeq 0\;\; {\rm and}\;\; p_{\,\parallel}\simeq 0.  \label{Emp-p2}
\end{equation}
The properties $|v|<1$ and  (\ref{Emp-p2}) mean that in fact the contribution of
$v p_{\,\parallel}$ into (\ref{Ekm-a}) is negligible small.
So keeping the same level of accuracy   for the approximation leading to
the relation (\ref{Emp-1}) for $E_{m}(\vec{p})$ as well as  for that one
for $p_{\,\parallel}$, which is required by conditions of consistency of
calculations, one has to approximate $ E_{m}(\vec{k}_{m}) $ and $k_{\,\parallel}$
given by (\ref{Ekm}), (\ref{km}) as
\begin{eqnarray}
E_{m}(\vec{k}_{m}) &\simeq& \gamma m, \label{Ekm-c} \\
k_{m\,\parallel} & \simeq & \gamma  v m, \label{km-c}
\end{eqnarray}
because within the approximation used in order to obtain (\ref{Emp-1})
the elements $ p_{\,\parallel}$  and $v p_{\,\parallel}$ are negligible small.

Note that the identity (\ref{gamma-p})
always holds when one considers Lorentz transformations of the energy $E(\vec{p})$ and momentum
$\vec{p}\;$ forming  4--vector $(E(\vec{p}),\vec{p})$ in Minkowski space and $ m > 0$. The same identity
has to hold for $E_{m}(\vec{k}_{m})$: There is
\begin{equation}
\gamma \equiv \frac{E_{m}(\vec{k}_{m})}{m}. \label{gamma-k}
\end{equation}
Using this identity and the approximation (\ref{Ekm-a}), which was
used in \cite{carlo} as the  basis of calculations performed therein,
we obtain the following consistency condition
\begin{equation}
\gamma \equiv \frac{E_{m}(\vec{k}_{m})}{m} \simeq \frac{\gamma (m + v \, p_{\,\parallel})}{m}. \label{gamma-2}
\end{equation}
The only solution of this equation is
\begin{equation}
v\,p_{\,\parallel} \simeq 0, \label{vp=0}
\end{equation}
from which it follows that, or $v \simeq 0$ and $p_{\parallel}$ is not too large,
or $v \simeq 0$ and $p_{\parallel} \simeq 0$ both. There is $\gamma \simeq 1$ in
these cases both. The third possibility is that $p_{\parallel} \simeq 0$ and $0< v < 1$.
This last case leads to some inconsistencies. Namely,
in any case the result (\ref{vp=0}) means that the relation
(\ref{Ekm-a}), (or  (33)  in \cite{carlo}), can not be considered
as a correct approximations. In the light of this discussion only the
relations (\ref{Ekm-c}) and (\ref{km-c}) are correct and self-consistent.
Thus
\begin{eqnarray}
E_{m}(\vec{k}_{m}) - E_{m'}(\vec{k}\,'_{m,}) &\simeq& \gamma (m - m'), \label{Ekm-b1}
\end{eqnarray}
and only  such  approximations (that is (\ref{Ekm-c}), (\ref{km-c}) and (\ref{Ekm-b1}))
should be used when performing integrations
in the integral (41) in \cite{carlo}
instead of  (\ref{Ekm-a})
(or  (33)  in \cite{carlo}).
Unfortunately,
this, contrary to the result (46) in \cite{carlo},  leads to the result,
$P_{v}(t) = P_{0}(\gamma t)$,
that is to the result (20) in \cite{carlo}, which was not observed in any experiment.

The use of  (\ref{sigma}) and (\ref{Emp-1})
in \cite{carlo}  is the indisputable fact and it  was crucial.
If to limit the analysis of results presented in \cite{carlo} to the
consequences of the relation (\ref{gamma=1})
being the consequence of (\ref{sigma}) and (\ref{Emp-1})
one may conclude that, in fact,
the relation (46) in \cite{carlo} (or (\ref{P=Pv}) in this letter),
can be realized only in the case of non moving particles, $\gamma \cong 1 $,
and it is approximately valid only for such particles.
This means that the main result  (46) of \cite{carlo}, which looks nice,
holds only in the trivial situation when the particle considered is not
moving (or is moving very slowly with an extremely  non--relativistic velocity).
On the other hand,
it can be seen that
the use of absolutely all implications of (\ref{sigma}) and (\ref{Emp-1}) without
missing (\ref{vp=0})
(and thus (\ref{Ekm-c}), (\ref{Ekm-b1})),
which is necessary to fulfill the consistency conditions, shows that
the result (46) in \cite{carlo} holds for $v \simeq 0$ and
it is wrong for such $v$ that $0< v < 1$: The correct approximation
(\ref{Ekm-c}) gives the relation (\ref{Ekm-b1}) and
leads to the result which was never
met in experiments and which is  different from (46) in \cite{carlo}.
So, in the light of the results  (\ref{gamma=1}) and
(\ref{vp=0}) ---  (\ref{Ekm-b1})
criticism of  the results  obtained
in \cite{stefanovich,stefanovich1,stefanovich2,shirkov,ku-plb}, which
are valid for any $\gamma > 1$, and conclusions presented
in \cite{carlo}  are completely unfounded. Nevertheless,
paradoxically, the publication \cite{carlo} is very important, because in fact it
confirms the observations and conclusions  drawn in
\cite{stefanovich,stefanovich1,stefanovich2,shirkov,ku-plb} that in
the case of moving particles the classical dilation relation of the form (\ref{P=Pv})
can be obtained within the rigorous quantum mechanical treatment of the
problem (without using assumptions of the type (\ref{Emp-1}))
only as the approximate  relation
but not as the exact relation.\\

\noindent
\textbf{Acknowledgments:}
The work was supported by the Polish  NCN grant No
DEC-2013/09/B/ST2/03455.

\end{document}